\documentclass[a4paper,12pt,dvipsnames,usernames]{article}
\pdfoutput=1

\usepackage{aas_macros}

\usepackage{jcappub} 

\usepackage[utf8]{inputenc}
\usepackage{graphicx}
\usepackage{amsmath}
\usepackage{amssymb}
\usepackage{bm}
\usepackage{paralist,array}
\usepackage{units} 
\graphicspath{{./Plots/}}
\usepackage{slashed}
\usepackage{todonotes}
\usepackage[parfill]{parskip}
\usepackage{longtable}
\usepackage{subfig}
\usepackage{float}

\graphicspath{{Plots/}}

\newcommand{\CNB}{\ensuremath{{\mathrm{C}\nu\mathrm{B}}}}

\title{Best-case scenarios for neutrino capture experiments}

\author[1,2,3]{Kyrylo Bondarenko}
 \emailAdd{kyrylo.bondarenko@cern.ch}
\affiliation[1]{
IFPU, Institute for Fundamental Physics of the Universe, via Beirut 2, I-34014 Trieste, Italy}
\affiliation[2]{
SISSA, via Bonomea 265, I-34132 Trieste, Italy
}
\affiliation[3]{
INFN, Sezione di Trieste, SISSA, Via Bonomea 265, 34136, Trieste, Italy
}

\author[4]{Alexey Boyarsky}
 \emailAdd{boyarsky@lorentz.leidenuniv.nl}
\affiliation[4]{%
 Institute Lorentz, Leiden University, Niels Bohrweg 2, Leiden, NL-2333 CA, the Netherlands
}

\author[5,6]{Josef Pradler}
\emailAdd{josef.pradler@oeaw.ac.at}
\affiliation[5]{Institute of High Energy Physics, Austrian Academy of Sciences, Georg-Coch-Platz 2, 1010 Vienna, Austria}
\affiliation[6]{University of Vienna, Faculty of Physics, Boltzmanngasse 5, A-1090 Vienna, Austria}

\author[7,8]{Anastasia Sokolenko}
\emailAdd{sokolenko@kicp.uchicago.edu}
\affiliation[7]{
Theoretical Astrophysics Department, Fermi National Accelerator Laboratory, Batavia, Illinois, 60510, USA
}
\affiliation[8]{
Kavli Institute for Cosmological Physics, The University of Chicago, Chicago, IL 60637, USA
}

\begin{document}
\vspace*{-1.5cm}
\abstract{A direct discovery of the cosmic neutrino background would bring to a closure the searches for relic left-over radiation predicted by the Hot Big Bang cosmology. Recently, the KATRIN experiment put a limit on the local relic neutrino overdensity  with respect to the cosmological predicted average value at $\eta \lesssim 10^{11}$ [\href{https://journals.aps.org/prl/abstract/10.1103/PhysRevLett.129.011806}{Phys.~Rev.~Lett.~129, 011806 (2022)}]. In this work, we  first examine to what extent such values of $\eta$ are conceivable. We show that even under cavalier assumptions, a cosmic origin of $\eta \gtrsim 10^4$ seems out of reach (with the caveat of forming bound objects under a new force,) but find that a hypothetical local source of low-energy neutrinos could achieve $\eta \sim 10^{11}$.  Second, when such values are considered, we point out that the experimental signature in KATRIN and other neutrino-capture experiments changes, contrary to what has hitherto been assumed. 
Our results are model-independent and maximally accommodating as they only assume the Pauli exclusion principle. 
As intermittent physics target in the quest for C$\nu$B detection, we identify an experimental sensitivity to $\eta \sim 10^4$ for which conceivable sources exist; to resolve the effect of a degenerate Fermi gas for such overdensity an energy resolution of 10~meV is required.
}

\maketitle

\section{Introduction}

The standard cosmological model predicts the existence of a primordially generated relic neutrino background with the average number density today of $n_0=56 \text{ cm}^{-3}$ per spin and flavor degree of freedom and six times that number in total~\cite{Mangano:2005cc,deSalas:2016ztq}. This number sets a lower bound on the abundance of neutrinos in the Universe if the Standard Model of particle physics and cosmology is valid. The primordial neutrino background plays an important role in the formation of the first elements and affects the timing of the relic photon decoupling, providing abundant indirect evidence of its existence, see~\cite{Fields:2019pfx, Pitrou:2018cgg, Planck:2018vyg,Burns:2022hkq} and references therein. However, because of the extremely small interaction cross section of low-energy neutrinos with ordinary matter, direct detection of the relic neutrino background remains an extraordinarily difficult experimental challenge. 

The direct experimental searches for neutrino mass serve a double purpose in that they are also sensitive to the relic neutrino background. The state-of-the-art experimental facility KATRIN recently put a limit on the local relic neutrino background overdensity $n < \eta n_0$ with $ \eta \simeq 10^{11}$~\cite{KATRIN:2022kkv},\footnote{Purely gravitational probes in our Solar system put bounds of the same order, see~\cite{Tsai:2022jnv}.} i.e.~many orders of magnitude away from the standard value.\footnote{Gravitational clustering of non-relativistic SM neutrinos is not expected to yield local numbers in excess of $\sim 20 n_0$~\cite{Ringwald:2004np,deSalas:2017wtt,Zhang:2017ljh,Mertsch:2019qjv}.} A next-generation neutrino mass experiment Project~8 is expected to have improved sensitivity although numbers have not yet been given~\cite{Project8:2022wqh}. Finally, there is an ambitious proposal PTOLEMY~\cite{PTOLEMY:2019hkd,Banerjee:2023lrk}, that aims to detect the relic neutrino background at $\eta = O(1)$.

The purpose of this paper is to explore to which extent current and future experiments can put sensible limits on a relic neutrino overdensity. Clearly, the neutrino number density \emph{can} be larger than in a standard cosmological history. Dark Matter or a fraction thereof may decay into neutrinos~\cite{McKeen:2018xyz,Chacko:2018uke,Nikolic:2020fom,Bondarenko:2020vta} and even Dark Energy has been considered as a source of additional present-day radiation degrees of freedom~\cite{Berghaus:2020ekh}.
Finally, an interesting possibility to enhance the local neutrino asymmetry by matter-interference has been recently noted in~\cite{Arvanitaki:2022oby,Arvanitaki:2023fij}.
However, even if such possibilities exist, the fermionic nature of neutrinos calls into question if previously considered overdensities of $\eta = O(10^{10})$ or even orders of magnitudes smaller are tenable in what is assumed to be low-energy (sub-eV) neutrino background.

In order to  provide a quantitative answer we shall take the most accommodating assumptions that are aimed at saturating the permissible number of cosmological, galactic, and local neutrino number densities.%
\footnote{The experimental rate is determined by the \emph{flux} of particles times the neutrino capture cross section $\Gamma \sim n v \times \sigma$. However, owing to the independence of the product $\sigma v$ to the relative velocity $v$ (for the range of energies considered), it suffices for us to simply discuss the maximum permissible $\emph{number}$ densities of neutrinos, irrespective of the neutrino velocity.}
For this purpose, the Pauli exclusion principle is our only guiding input. 
The success of using such general considerations such as Pauli-blocking and/or the Liouville theorem is best exemplified by the prominence of the Gunn-Tremaine bound~\cite{Tremaine:1979we}; see also~\cite{Boyarsky:2008ju,Alvey:2020xsk} for a more recent application.
We find that under such agnostic and model-independent assumptions, $\eta \sim 10^{11}$ is attainable  only locally in the solar neighborhood, through largely hypothetical and presumably not very credible sources. Galactically and extragalactically sourced number densities are limited to $\eta\lesssim 10^4$. Although much smaller, the latter value can nevertheless serve as a clear target. In addition, we show that the present KATRIN analysis is in fact sensitive to the local neutrino phase space distribution as the endpoint region of the beta spectrum is subject to Pauli blocking effects.

The  paper is organized as follows: in Sec.~\ref{sec:capture} we discuss the physics of the neutrino capture process and our simplified treatment of electron neutrinos. In Sec.~\ref{sec:max_number_density} we discuss model-independent upper limits on the neutrino number density, both cosmological and local.
In Sec.~\ref{sec:KATRIN} we discuss experimental signatures for the KATRIN-like experiments. Finally, in Section~\ref{sec:conclusions} we summarize our results and conclude.

\section{Neutrino capture}
\label{sec:capture}

Before we start the stock-taking of possible neutrino flux contributions, here we recall the generalities of neutrino capture detection through the electron flavor $\nu_e$. For the purpose of this paper, we shall consider only incoherent ensembles of neutrino mass eigenstates and neglect oscillations altogether. This is certainly a perfect assumption for cosmological fluxes but may lead to $O(1)$ changes for local fluxes.

Neutrinos propagate as mass-eigenstates $\nu_i$ and enter as such the detector in an incoherent mixture of flavor states as they have traveled astronomical distances from the source. For tritium-based detectors, neutrinos in mass-eigentstates $\nu_i$ will undergo the reaction  $\nu_e + \rm T \to \, ^3{\rm He}+ e^-$ according to their electron-flavor content $|U_{ei}|^2$. The capture rate in a detector of total target mass $M_{\rm T}$  is then given by
\begin{equation}
    \label{eq:GammanuDM}
    \Gamma =  
    \frac{M_{\rm T}}{m_{\text{T}}}
    \sum_i|U_{ei}|^2 \int dE_{\nu,i}\,  (\sigma \, v_{\nu,i}) \frac{dn_{\nu,i}}{dE_{\nu_i}} 
    \approx 
     \frac{M_{\rm T}}{m_{\text{T}}}
     (\sigma v)_0 \sum_{i=1}^{3}|U_{ei}|^2 n_{\nu,i}.
   \end{equation}
Here, ${dn_{\nu,i}}/{dE_{\nu_i}}$ is the (total) energy differential number density of $\nu_i$,  $m_{\text{T}}$ is the mass of one tritium atom, and $v_{\nu,i}$ is the neutrino velocity in the detector frame. The last relation follows on the account of velocity-independence of $(\sigma \, v_{\nu,i}) $. For a tritium target $(\sigma v)_0 \approx 3.8 \times 10^{-45}\ \text{ cm}^2$ ~\cite{Long:2014zva},%
\footnote{More precisely, left-helical neutrino states interact with strength $(1+v_{\nu,i})$ while right-helical states are weighted by a factor of $ (1-v_{\nu,i}) $ in the cross section~\cite{Long:2014zva}. This leads to a factor of two differences in the rate depending on the occupation number of these states and their kinematic regime. We shall ignore this model-dependence in~\eqref{eq:GammanuDM} because it does not change the overall argument.}
 independent of neutrino energy for $E_{\nu,i} \lesssim 10\ {\rm keV}$.
It  will be the working hypothesis of this paper that we saturate the neutrino densities at the lowest energies, either relativistically or non-relativistically. Hence, the local number density alone determines the overall rate.

For better orientation, we recall the electron flavor content of SM neutrinos:  $ |U_{e1}|^2 \simeq 0.68 $, $ |U_{e2}|^2 \simeq 0.30 $, $ |U_{e3}|^2 \simeq 0.02 $~\cite{ParticleDataGroup:2022pth}; in the absence of further sterile states with $i>3$, $U$ becomes the PMNS matrix. More generally, reserving $i=1,2,3$ for SM neutrinos and $i>3$ for any other light sterile states we may discriminate two principal scenarios,
\begin{align}
\text{active } \nu \text{ dominated rate:}& \quad \sum_{i=1}^3 |U_{ei}|^2 n_i \gg \sum_{i>3} |U_{ei}|^2 n_i  ,\\
\text{sterile }\nu\text{ dominated rate:} & \quad \sum_{i=1}^3 |U_{ei}|^2 n_i \ll \sum_{i>3} |U_{ei}|^2 n_i  .
\end{align}
Unitarity of $U$ imposes the constraint $\sum_i |U_{ei}|^2 = 1 $. The experimentally established non-violation of unitarity of the PMNS matrix puts constraints on the entries $U_{ei}|_{i>3}$ and they are generally found to be at the percent or permille level~\cite{Blennow:2016jkn,Blennow:2023mqx}. Therefore, a ``sterile dominated rate'' requires a significant excess over SM states. 

For the purpose of this paper, we shall assume that the capture rate is dominated by active neutrinos. Since we deal with numbers that span many orders of magnitude we make the identification 
\begin{align}
\label{effectivene}
 n_{\nu_e}  \equiv \sum_{i=1}^3 |U_{ei}|^2 n_i  , 
\end{align}
where $n_{\nu_e}$ stands as a proxy for the electron neutrino number density that will be informative of the overall achievable rate. 

\section{Census of maximum neutrino occupation numbers}
\label{sec:max_number_density}

As discussed above, neutrino capture experiments are only sensitive to the electron neutrino number density. Using the simplified ``one-flavor model''~\eqref{effectivene}  we shall write $m_{\nu_e}$ for the mass of the neutrino-eigenstate that dominates in electron flavor and assume it to constitute one degree of freedom, $g_{\nu_e}=1$.

Neutrinos obey the Pauli exclusion principle. It means that their number density is constrained by,
\begin{equation}
\label{eq:pauli}
    n_{{\nu_e}} \le \frac{V_p}{(2 \pi)^3},
\end{equation}
where $V_p = \frac{4 \pi}{3} p_{{\nu_e}, \max}^3$ is a spherical volume in momentum space with a maximal momentum equal to $p_{{\nu_e}, \max}$. The inequality is saturated for a fully degenerate neutrino gas for which $ p_{{\nu_e}, \max} = k_F$ with $k_F$ being the Fermi momentum. In what follows we shall consider the permissible values of $p_{{\nu_e}, \max}$, i.e., seek the maximal occupation numbers.

\subsection{Maximal cosmological density}
\label{sec:cosmonu}

The number of relativistic degrees of freedom in the early Universe prior to CMB decoupling is well constrained, implying that any significant overabundance of neutrinos $\eta \gg 1$ must have formed at a later epoch.  Candidate sources are processes that transfer mass and energy density from the DM and Dark Energy (DE) reservoirs. Indeed, the late-time cosmic energy budget remains in place for as long as the product $\langle E_\nu \rangle n_\nu \ll \rho_c$ where $\langle E_\nu \rangle$ is the typical energy of extra neutrinos and  $\rho_c = 3 H_0^2 M_P^2 $ is the critical energy density. This holds on the account of the similarity of $\Omega_m$ and $\Omega_\Lambda$; $M_P$ is the reduced Planck mass.

Equating the neutrino energy density to the critical density saturates the maximum permissible cosmological number density of neutrinos today and gives us as figure of merit
$     \eta < 10^2\  {\rm eV}/{\langle E_\nu \rangle }$.
Of course, $\eta$ cannot be amplified arbitrarily by considering non-relativistic neutrinos of minuscule mass. Considering a Fermi gas at zero temperature and taking the massless limit as the most accommodating case, the condition $\rho_{\nu_e} < \rho_{c}$ yields the Fermi momentum 
\begin{equation}
    k_F < \sqrt[4]{8\pi^2 \rho_{c}} \approx 7.5\text{ meV}. 
\end{equation}
From~\eqref{eq:pauli} with $ p_{{\nu_e}, \max} = k_F $ we may conclude $\eta \lesssim 2 \times 10^4$ as the maximal permissible cosmological overdensity.

The most readily conceivable scenario is that light bosonic DM particles, sufficiently populated in number, decay into neutrinos. Gravitational probes of DM decay constrain the dominant DM lifetime $\tau_{\rm DM}\gtrsim 35 t_{\text{U}}$ where $t_{\text{U}}$ is the age of the Universe~\cite{DES:2020mpv};
see also~\cite{Enqvist:2015ara,Nygaard:2020sow,Simon:2022ftd}. In the limit of $\tau_{\rm DM} \gg t_{\text{U}}$ the estimate on the number density is $n_{\nu_e} \approx N_\nu \rho_c \Omega_{\rm DM}t_{\text{U}}/(m_{\rm DM}\tau_{\rm DM})$ where $m_{DM}$ is the DM mass and $N_\nu$ is the effective multiplicity of neutrinos in the decay. This translates into $\eta \approx N_\nu (35 t_{\text{U}}/\tau_{\rm DM}) ({\rm eV}/m_{\rm DM})$ which shows that the maximum cosmological density is not readily achieved through DM decay (unless one optimizes DM mass and neutrino multiplicity in the decay.)

A scenario that may actually achieve $\eta \gg 1$ has been considered in~\cite{Berghaus:2020ekh}: a rolling scalar dark energy field generates different forms of dark radiation through friction, which may ultimately result in a sizable population of active neutrinos. Neutrino capture experiments hence offer an attractive pathway for probing such dark energy dynamics.

\subsection{Maximal galactic neutrino number density}

In the previous section, we have estimated the maximal coarse-grained cosmological electron neutrino number density. However, Earth is located in the overdensity created by the gravitational field of our Galaxy and surrounded by a dark matter halo. In this section, we will estimate the maximal electron neutrino density in the Galaxy from general principles, without assuming a specific dark matter density profile and neutrino production mechanism.

\subsubsection{Accumulation of neutrinos}

Let us consider neutrinos that accumulate in our Galaxy because of some process, e.g.~dark matter decay into neutrinos, such that the decay products remain gravitationally bounded. Such neutrinos necessarily need to be non-relativistic and hence we may take $p_{{\nu_e}, \max} = m_{{\nu_e}} v_{\text{esc}}$, where $v_{\text{esc}}$ is the escape velocity in our Galaxy. Using Eq.~\eqref{eq:pauli} this results in (see also~\cite{Bauer:2022lri})
\begin{equation}
  \eta \le  240   \left(\frac{m_{{\nu_e}}}{1~\text{eV}}\right)^3 \left(\frac{v_{\text{esc}}}{550\text{ km}/\text{s}}\right)^3,
    \label{eq:nu-accum}
\end{equation}
where for the estimate we take the neutrino mass on the order of our current laboratory constraints~\cite{Aker:2021gma} and used a typical value for the escape velocity of our galaxy~\cite{2017MNRAS.468.2359W,2018A&A...616L...9M,2019MNRAS.485.3514D,Necib:2021vxr}. 

It should be noted that such overdensities are in principle easily achievable. Comparing the implied neutrino energy density to the local DM density  $\rho_0 \simeq 0.3\ {\rm GeV/cm^3}$ yields $\rho_\nu /\rho_0 \sim 10^{-5} ({m_{{\nu_e}}}/{1~\text{eV}})^4$. In other words, even if DM decays with a lifetime much in excess of the age of the Galaxy or Universe,~\eqref{eq:nu-accum} can be reached. The (severe) fine-tuning is in the kinematic decay conditions as such neutrinos would need to be injected non-relativistically below the escape speed.

\subsubsection{Escaping neutrinos}

Let us now consider the case when neutrinos are created in our Galaxy but carry enough kinetic energy to leave the galactic gravitational potential. Taking again the most accommodating position to arrive at a figure of merit, one may saturate the number by assuming that  escaping neutrinos at the Sun's orbital radius in the Galaxy should not carry away energy larger than the total enclosed mass within this radius,
\begin{equation}
    \langle E_{{\nu_e},\text{esc}}\rangle \frac{d N_{{\nu_e},\text{prod}}}{dt} t_{\text{U}} < M_{\text{encl}},
\end{equation}
where $\langle E_{{\nu_e},\text{esc}}\rangle$ is the average escaping neutrino energy, $d N_{{\nu_e},\text{prod}}/dt$ is neutrino production rate, that we assume constant, $t_{\text{U}}$ is the age of the Universe, and $M_{\text{encl}}$ is an enclosed mass within Sun's radius.

The neutrino production rate can be related to the escaping neutrino number density $n_{{\nu_e},\text{esc}}$ at the Sun's orbit radius from the condition that the total production rate of neutrino should be equal to the total escape rate,
\begin{equation}
    \frac{d N_{{\nu_e},\text{prod}}}{dt} = \frac{d N_{{\nu_e},\text{esc}}}{dt} \sim 4\pi R_{\text{Sun}}^2 n_{{\nu_e},\text{esc}} v_{{\nu_e},\text{esc}},
\end{equation}
where $R_{\text{Sun}} = 8.3$~kpc is the distance from Sun to the galactic center. This gives a constraint on the escaping neutrino number density,
\begin{equation}
    n_{{\nu_e},\text{esc}} < \frac{M_{\text{encl}}}{4\pi R_{\text{Sun}}^2 t_{\text{U}} \langle E_{{\nu_e},\text{esc}}\rangle  v_{{\nu_e},\text{esc}}} = \frac{V_{\text{Sun}}^2}{4\pi G R_{\text{Sun}} t_{\text{U}} \langle p_{{\nu_e},\text{esc}}\rangle},
    \label{eq:Mgal}
\end{equation}
where in the last equality we used relation $p_{{\nu_e}} = v_{{\nu_e}} E_{{\nu_e}}$, that works for both relativistic and non-relativistic particles, and the enclosed mass is related with  the Sun's circular velocity as $V_{\text{Sun}} \simeq 240$~km/s~\cite{2021MNRAS.502.4377B} as $M_{\text{encl}} = V_{\text{Sun}}^2 R_{\text{Sun}}/G$. This constraint can be released for the low average neutrino momentum $\langle p_{{\nu_e},\text{esc}}\rangle$. 

For low neutrino momentum, we must  account for Pauli blocking. 
For the sake of the argument, we may  take that half of the escaping neutrinos have momenta smaller than the average one. 
Hence, even if all the quantum states below the average momentum are occupied, we may adopt as an accommodating constraint on the escaping neutrino number density [see Eq.~\eqref{eq:pauli}],
\begin{equation}
    \frac{n_{{\nu_e},\text{esc}}}{2} < \frac{\langle p_{{\nu_e},\text{esc}}\rangle^3}{6 \pi^2}.
    \label{eq:dNtodT}
\end{equation}

Comparing~\eqref{eq:Mgal} and~\eqref{eq:dNtodT} we see that there is an optimal value of the average neutrino momentum $\langle p_{{\nu_e},\text{esc}}^{\text{opt}}\rangle$, for which the electron neutrino number density is maximized,
\begin{equation}
    \langle p_{{\nu_e},\text{esc}}^{\text{opt}}\rangle = \left(\frac{3 \pi V_{\text{Sun}}^2}{4 G R_{\text{Sun}} t_{\text{U}}}\right)^{1/4} \approx 4\text{ meV} ,
\end{equation}
which gives the final constraint on the number density of escaping neutrino
\begin{equation}
   \eta < \frac{1}{3\pi^2 n_0} \left(\frac{3 \pi V_{\text{Sun}}^2}{4 G R_{\text{Sun}} t_{\text{U}}}\right)^{3/4} \approx 5\times 10^{3}  .
    \label{eq:nu-esc}
\end{equation}
We see that the maximal conceivable neutrino overdensity of galactic origin is in the same ballpark as the cosmological one obtained in Sec.~\ref{sec:cosmonu}. A more detailed discussion on accumulating and escaping neutrino fluxes that originate from DM decay in the galaxy can be found in~\cite{Nikolic:2020fom}.

\begin{figure}[t]
    \centering
    \includegraphics[width=0.6\linewidth]{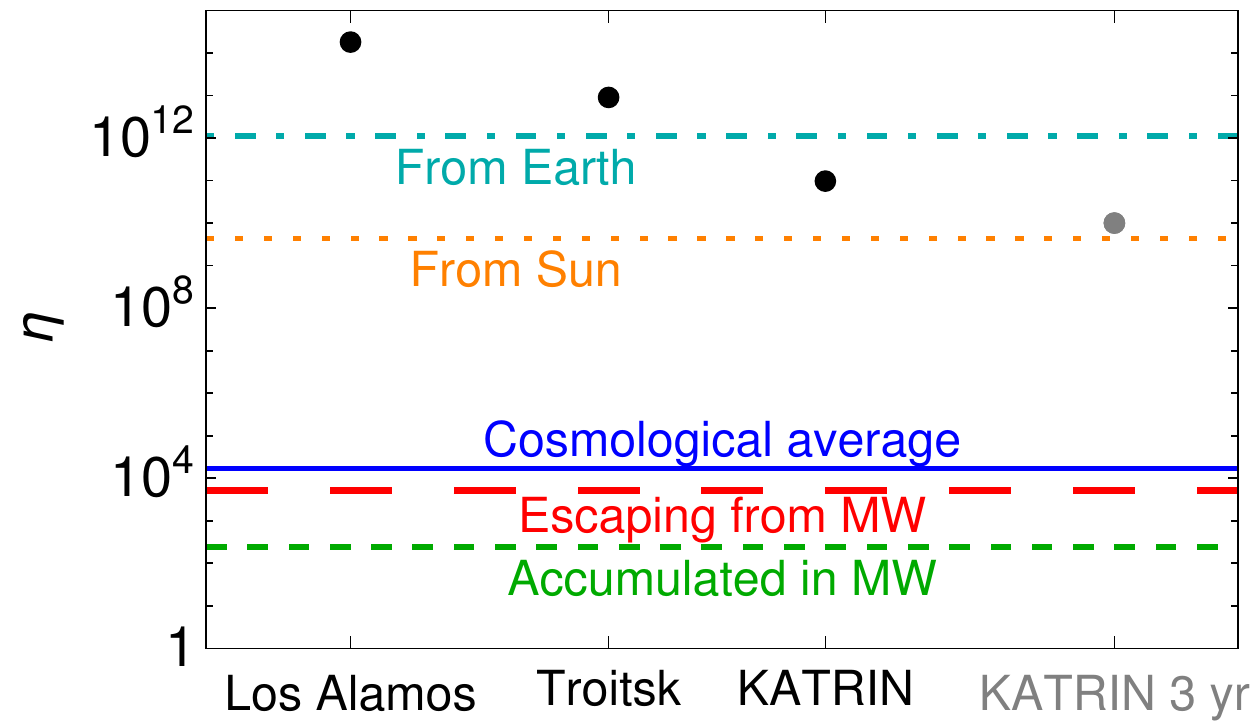}
    \caption{The maximal conceivable overdensities from different sources relative to the relic neutrino background number density. The blue line shows the maximal cosmological average value (see section~\ref{sec:cosmonu}), green dashed line shows the maximal neutrino number density that are gravitationally bounded in our Galaxy for $m_{\nu_e}= 1$~eV (lower mass corresponds to even lower maximal number density, see Eq.~\eqref{eq:nu-accum}). Red long-dashed, orange dotted, and cyan dashed-dotted lines show the maximal neutrino density that can be reached by a constant neutrino flux generated by our Galaxy (Eq.~\eqref{eq:nu-esc}), Sun (Eq.~\eqref{eq:nu-esc-Sun}) and Earth (Eq.~\eqref{eq:nu-esc-Earth}), respectively. The black dots are current constraints from Los Alamos~\cite{Robertson:1991vn}, Troitsk~\cite{Lobashev:1999dv}, and KATRIN~\cite{KATRIN:2022kkv} and the gray dot is the 3 year projection from KATRIN~\cite{KATRIN:2022kkv}.}
    \label{fig:max_density}
\end{figure}

\subsection{Maximal Solar System density}

The analysis of the previous section for the Galaxy can be applied to any gravitationally bound system, including the Solar System and Earth. Let us consider the (hypothetical) case when neutrinos are massively produced in the Sun or inside Earth. Since the escape velocity from Earth or from the Solar System at Earth's position is an order of magnitude smaller than the escape velocity from our Galaxy, we need not consider the accumulation of neutrino within the Solar System; see~\eqref{eq:nu-accum} above.%
\footnote{See~\cite{Lasenby:2020goo} for a treatment of accumulating particle populations in the solar neighborhood; in~\cite{Nikolic:2020fom} those ideas have been applied to the solar neutrino flux with negligible resulting overdensities.}

For escaping neutrinos the situation is markedly  different. From Eq.~\eqref{eq:nu-esc} we see that the bound scales as $V_c^2/R$, where $V_c$ is a circular velocity of the orbit with radius $R$. This is just the gravitational acceleration at the point where the experiment is located. Of course, the gravitational acceleration from Sun or from Earth is orders of magnitude stronger than from the Galaxy, hence in principle allowing for significantly larger values of~$\eta$.

Using well-known parameters for Sun and Earth we obtain the following limits on the electron neutrino number density in the laboratory as
\begin{equation}
    \langle p_{{\nu_e},\text{esc}}^{\text{opt,Sun}}\rangle = \left(\frac{3 \pi g_{\text{Sun}}}{4 G t_{\text{Sun}}}\right)^{1/4} \approx 380\text{ meV} \quad
    \Rightarrow \quad \eta \lesssim 4\times 10^9. 
    \label{eq:nu-esc-Sun}
\end{equation}
for neutrinos produced in the Sun, and 
\begin{equation}
    \langle p_{{\nu_e},\text{esc}}^{\text{opt,Earth}}\rangle = \left(\frac{3 \pi g_{\text{Earth}}}{4 G t_{\text{Earth}}}\right)^{1/4} \approx 2.4\text{ eV}
    \quad
    \Rightarrow \quad \eta \lesssim  10^{12}. 
    \label{eq:nu-esc-Earth}
\end{equation}
for neutrinos produced in the Earth. Here $t_{\text{Sun}} \approx t_{\text{Earth}} = 4.6\times 10^9$~yr are the ages of Earth and Sun, while $g_{\text{Sun}}$ and $g_{\text{Earth}}$ are gravitational laboratory acceleration due to the Sun and Earth's masses, respectively. 

Let us pause and recall again that the derivation of those extremely large values of $\eta$ assumed an injection rate that implies that a fair fraction of Earth and/or Sun mass has already been converted into neutrinos (under the assumption of a constant rate) and hence relies on truly cataclysmic assumptions that are, mildly put, observationally challenged. This puts into perspective the constraint $\eta \lesssim 10^{11}$ derived by KATRIN~\cite{KATRIN:2022kkv} and other preceding works~\cite{Robertson:1991vn, Lobashev:1999dv}.

\subsection{Neutrino clusters}

We close our census of neutrino occupation number by mentioning the possibility of bound objects that we may term neutrino clusters. A neutrino gas in equilibrium, under the influence of gravity and supported by degeneracy pressure will have a size of $R\sim R_{\rm NS} (m_n/m_{\nu_e})^2 \approx  (0.5\ {\rm eV}/m_{\nu_e})^2 {\rm Mpc}$ and total mass $ M\sim M_{\rm NS} (m_n/m_{\nu_e})^2 $,
where $R_{\rm NS}$ ($M_{\rm NS}$) is the radius (mass) of a neutron star and $m_n$ is the neutron mass~\cite{Markov:1964fjm}. Such an object is of cosmic span, outweighs the most massive galaxy clusters, and has a central mass density that exceeds $\rho_c$ by a factor $10^4 (m_{\nu_e}/0.1\ {\rm eV})$. At face value, this translates into a central overdensity of
\begin{align}
    \eta \approx \frac{\rho_{\rm NS}(0)}{n_0m_n} \left(\frac{m_{\nu_e}}{m_n}\right)^3 \approx 6\times 10^6 \left(\frac{m_{\nu_e}}{0.1\ {\rm eV}}\right)^3 ,
\end{align}
but it is hard to see how such an object could have come into existence in a $\Lambda$CDM Universe and not be ruled out observationally through its gravitational effect~\cite{Hotinli:2023scz}.  

An alternative avenue to reduce the size and mass of a neutrino cluster is to consider a new Yukawa force that binds neutrinos stronger than gravity. This possibility has been studied in detail in~\cite{Smirnov:2022sfo} for both, non-relativistic and relativistic bound objects. A maximal overdensity $\eta < 10^7 (m_{\nu_e}/0.1\ {\rm eV})^3$
and minimal radius of $R \gtrsim 0.6\ {\rm km} (10^{-7}/y)(0.1\ {\rm eV}/ m_{\nu_e}) $ are found; $y$ is the Yukawa coupling.
Such clusters could form by fragmentation of the C$\nu$B at some redshift $z_{\rm f}\lesssim 200$ yielding an overdensity today of $\eta \sim z_{\rm f}^3$. The detection prospects will depend on the size of the clusters: their effect may average out or, if situated in a large enough (solar-system sized) structure, an enhancement of the signal rate, in the best of cases by~$\eta\sim 10^7$ is possible. For a detailed account, we refer the reader to~\cite{Smirnov:2022sfo}; see also~\cite{Wise:2014ola}.

\begin{figure}
    \centering
    \includegraphics[width=0.48\linewidth]{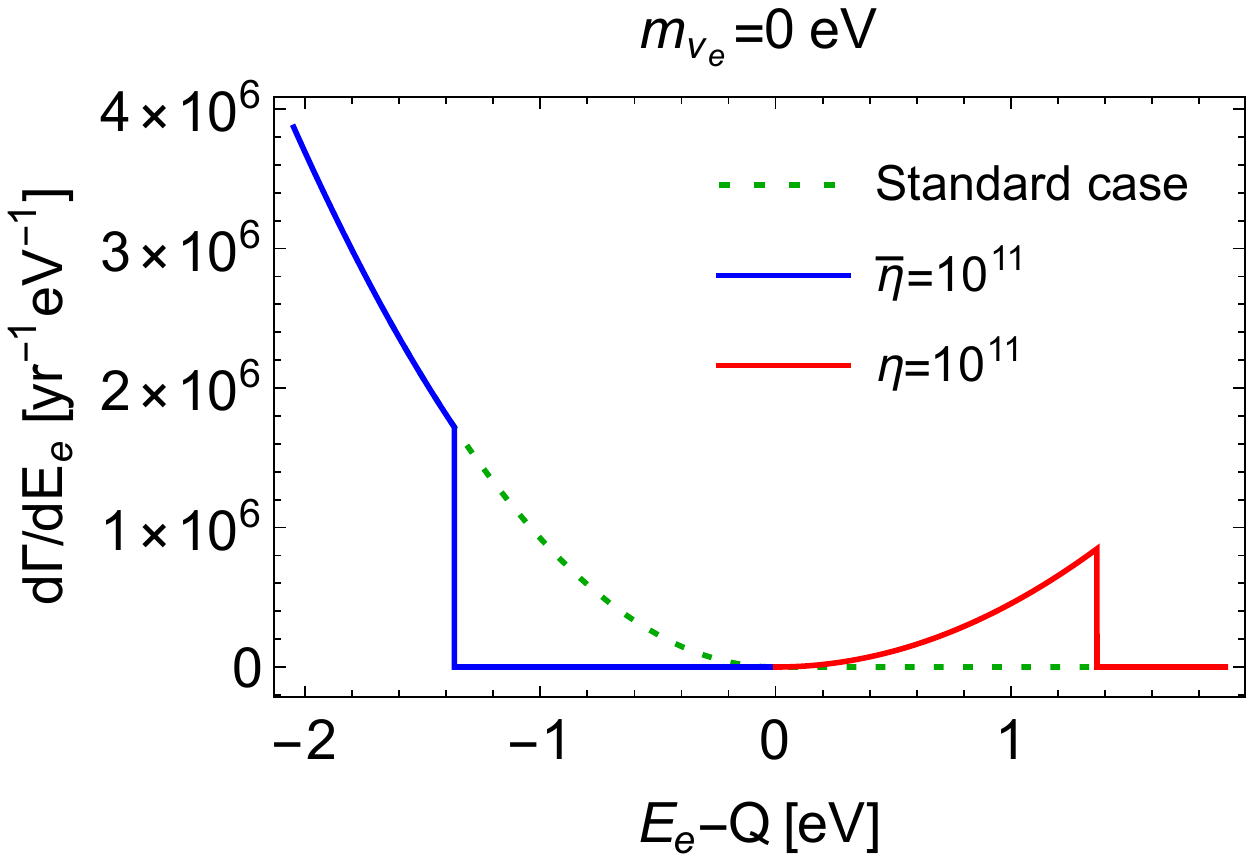}~\includegraphics[width=0.48\linewidth]{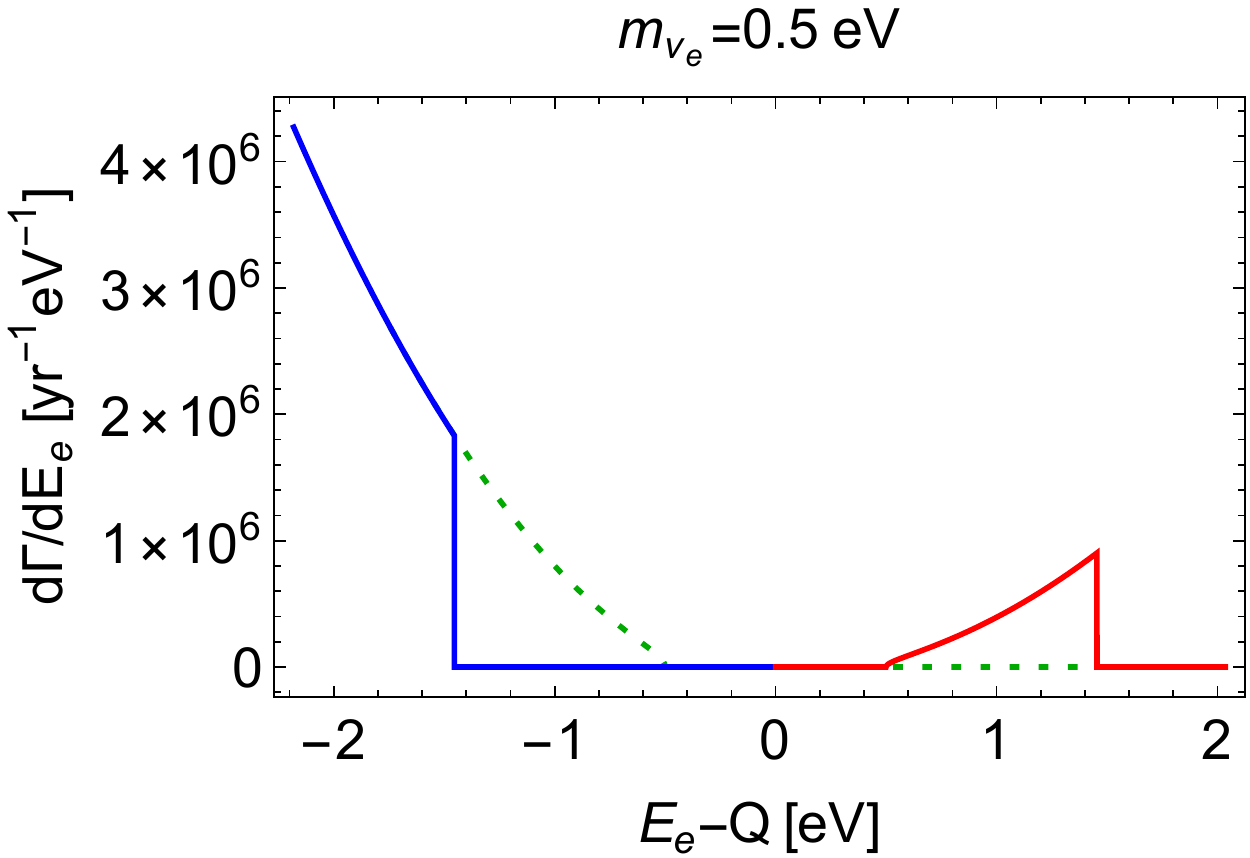}
    \caption{Differential electron emission rate for a tritium mass of $100 \mu$g as a function of the electron kinetic energy $E_e$ measured relative to the beta-decay endpoint energy~$Q$; an illustration of the signatures of the large neutrino and antineutrino overdensities near Earth for massless neutrino (left panel) and neutrino mass $m_{\nu_e}=0.5$~eV (right panel). The green dotted line shows the prediction for the standard tail of the beta-decay, the blue line is its modification due to the presence of antineutrinos with an overdensity $\bar{\eta} = 10^{11}$, and the red line is a signal for the neutrino capture rate for a neutrino overdensity $\eta = 10^{11}$. We assume that both neutrinos and antineutrinos are distributed as a Fermi gas at zero temperature.}
    \label{fig:peak_valley} 
\end{figure}

\section{Detection of a local low-energy neutrino background}
\label{sec:KATRIN}

\subsection{Different signatures of neutrino background}

\begin{figure}
    \centering
    \includegraphics[width=0.48\linewidth]{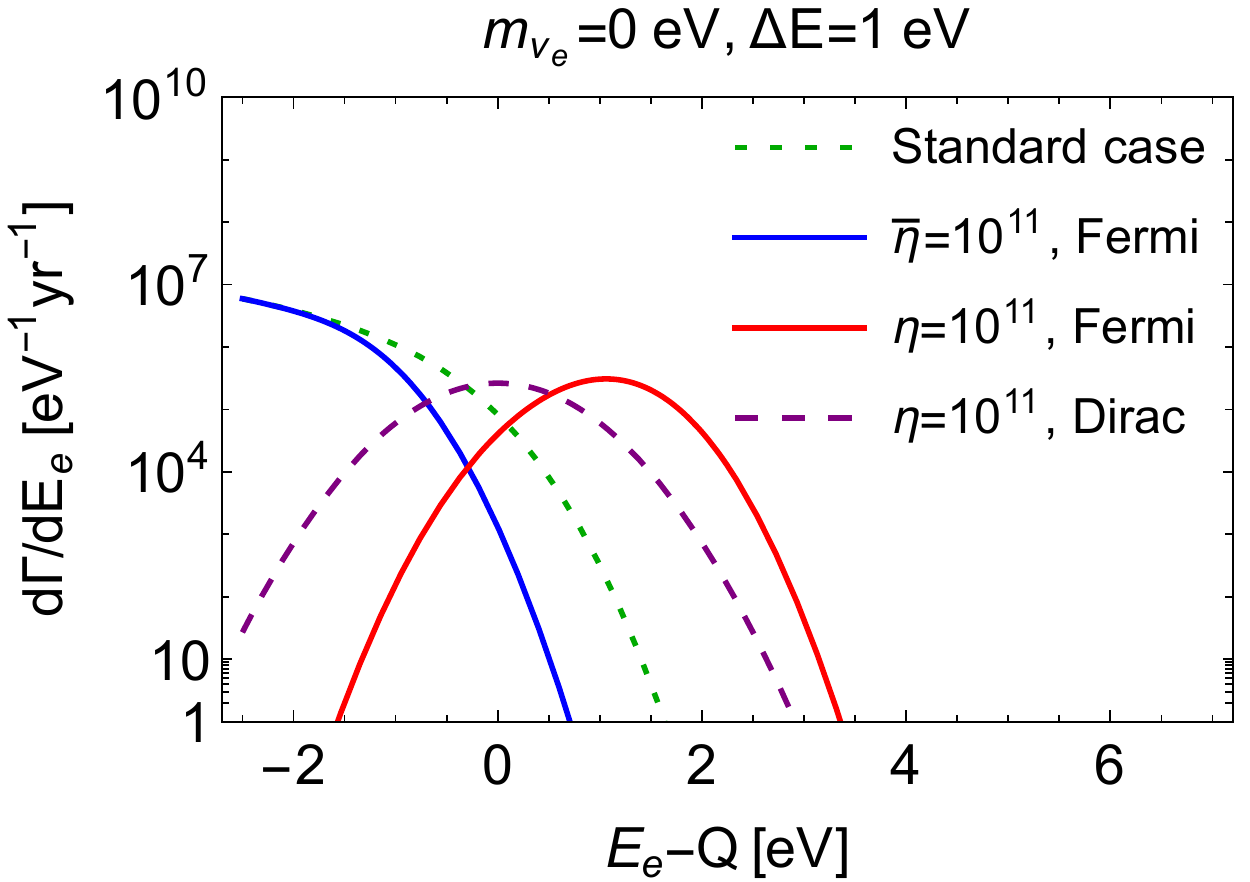}~\includegraphics[width=0.48\linewidth]{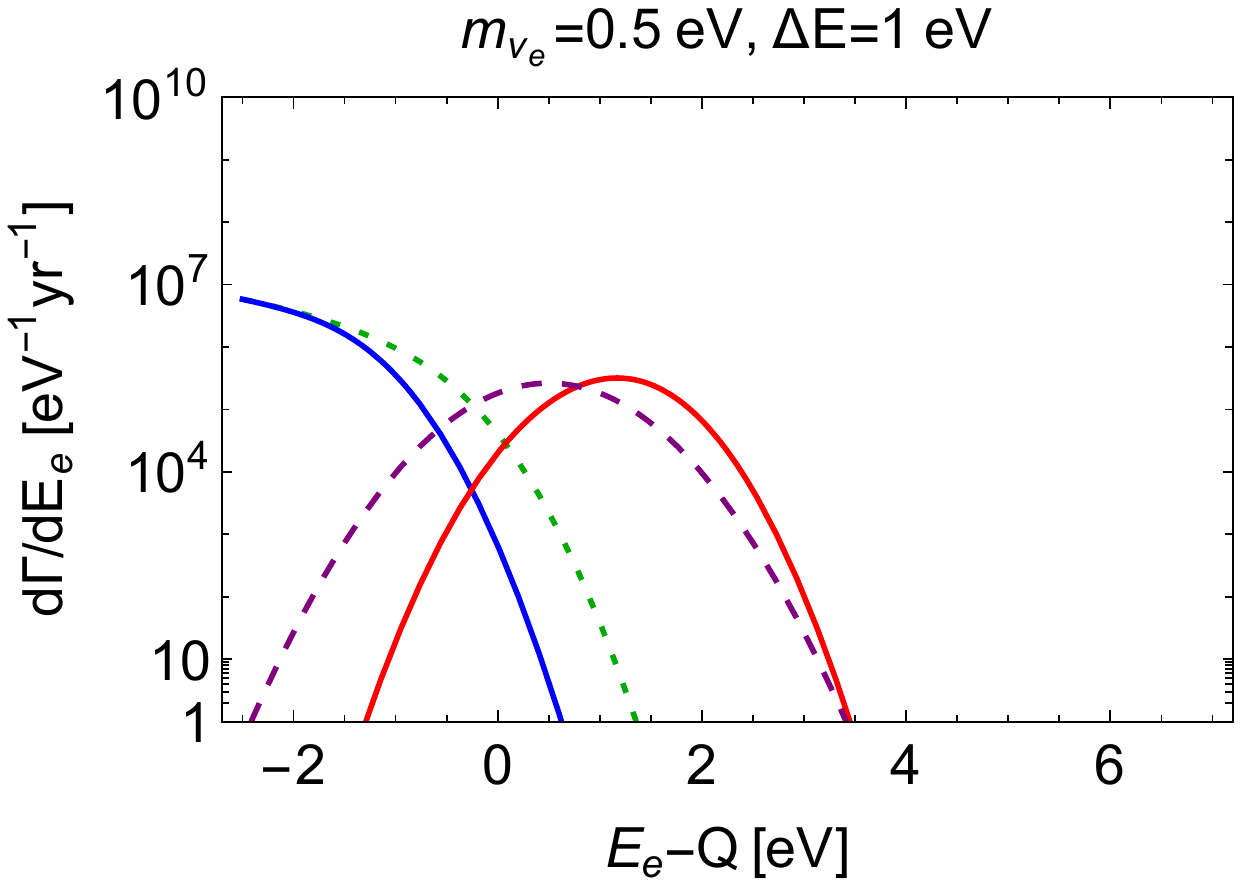}
    \\ \vspace{0.1cm}
    \includegraphics[width=0.48\linewidth]{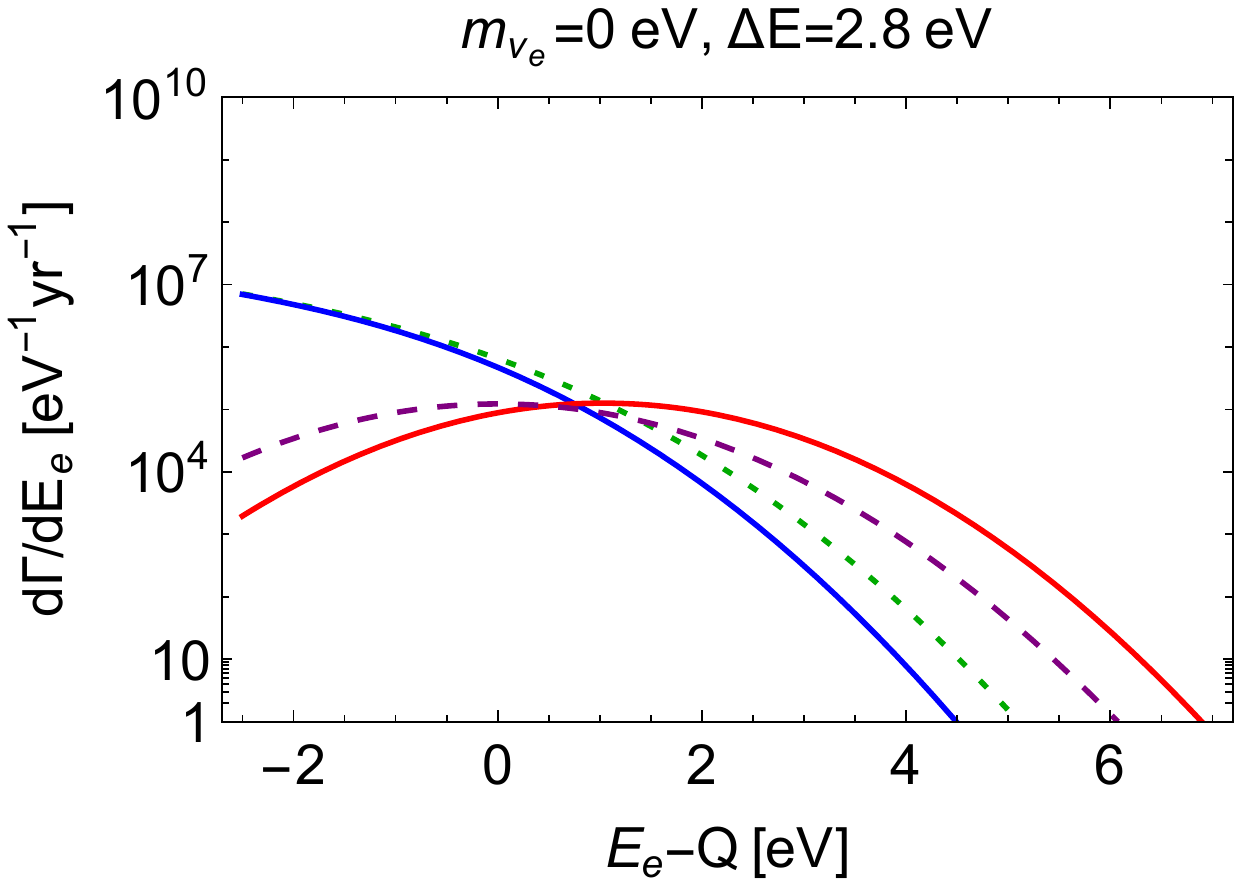}~\includegraphics[width=0.48\linewidth]{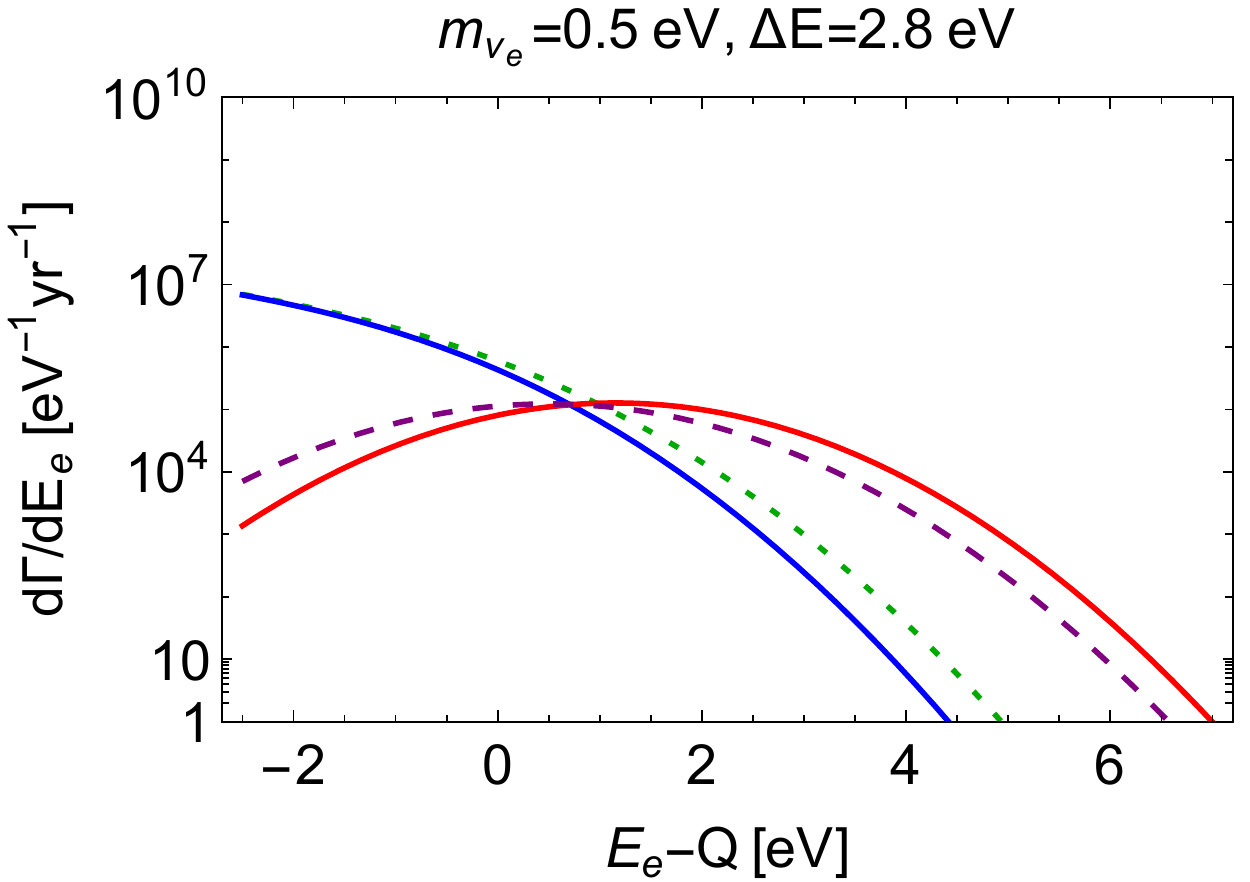}
    \caption{Differential electron rate with applied energy resolution for a tritium mass $100 \mu$g versus electron kinetic energy $E_e$ for massless neutrinos (left panels) and massive neutrino of mass $m_{\nu_e}=0.5$~eV (right panels). In the upper panels we adopt an energy uncertainty $\Delta E = 1$~eV, while for lower panels we chose $\Delta E = 2.8$~eV.
    The green dotted line shows the standard tail of the beta-decay, the blue line is its modification due to the presence of antineutrinos with the overdensity $\bar{\eta} = 10^{11}$ for the Fermi sphere distribution. The red line is a signal from the neutrino capture rate for overdensity $\eta = 10^{11}$ for a Fermi sphere, while the purple dashed line shows neutrino capture rate in the assumption that all neutrino momenta are negligibly small (Dirac delta function in momentum space, as it was assumed in~\cite{KATRIN:2022kkv}).}
    \label{fig:smearing} 
\end{figure}

In previous sections, we discussed situations $\eta \gg 1$ in which the local neutrino number density can be much larger than the average one. In this section we discuss the observational signatures for a neutrino-capture capable experiment such as KATRIN or PTOLEMY, ignoring backgrounds and/or potential systematics. We focus on tritium as a target; results are easily generalized to other candidate targets.

\begin{figure}
    \centering
    \includegraphics[width=0.48\linewidth]{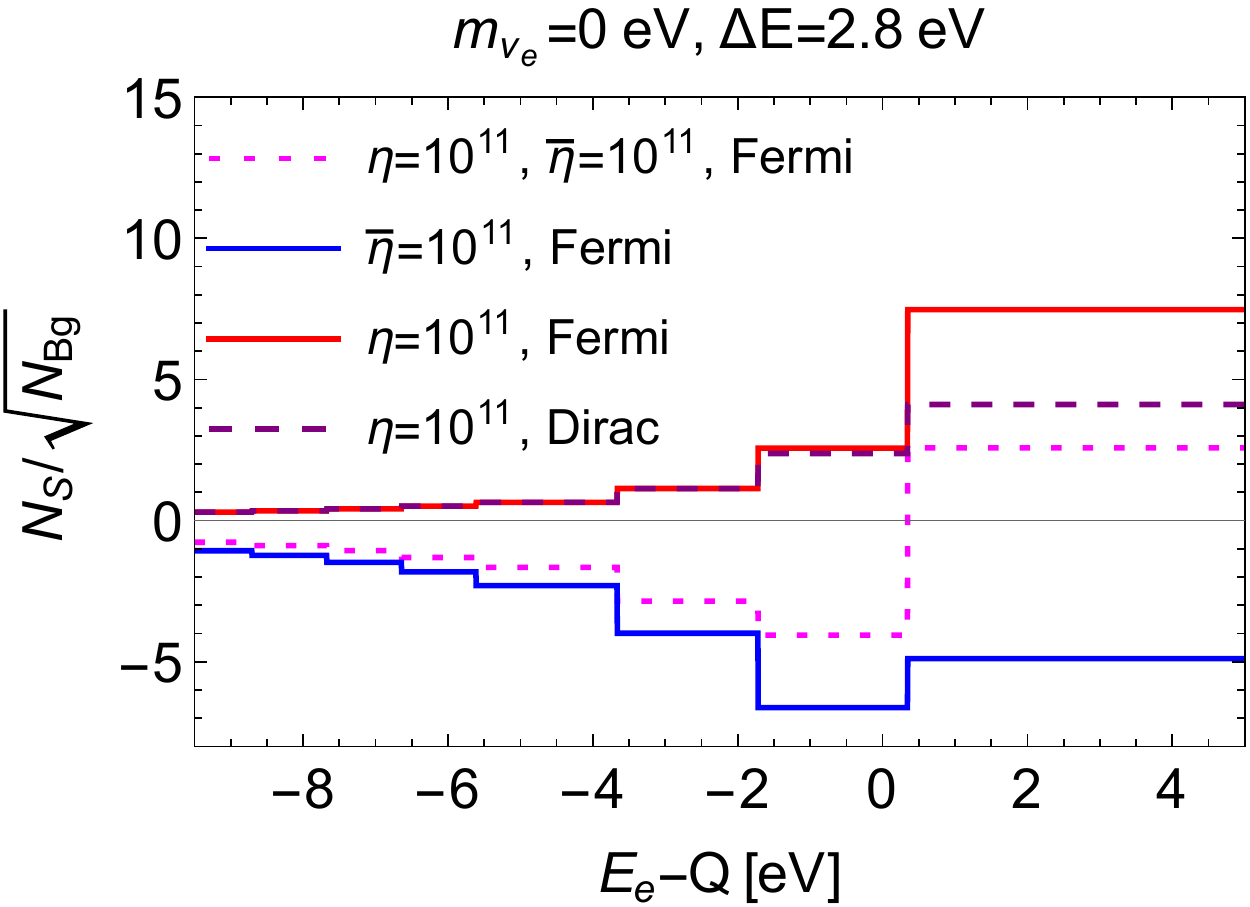}~\includegraphics[width=0.48\linewidth]{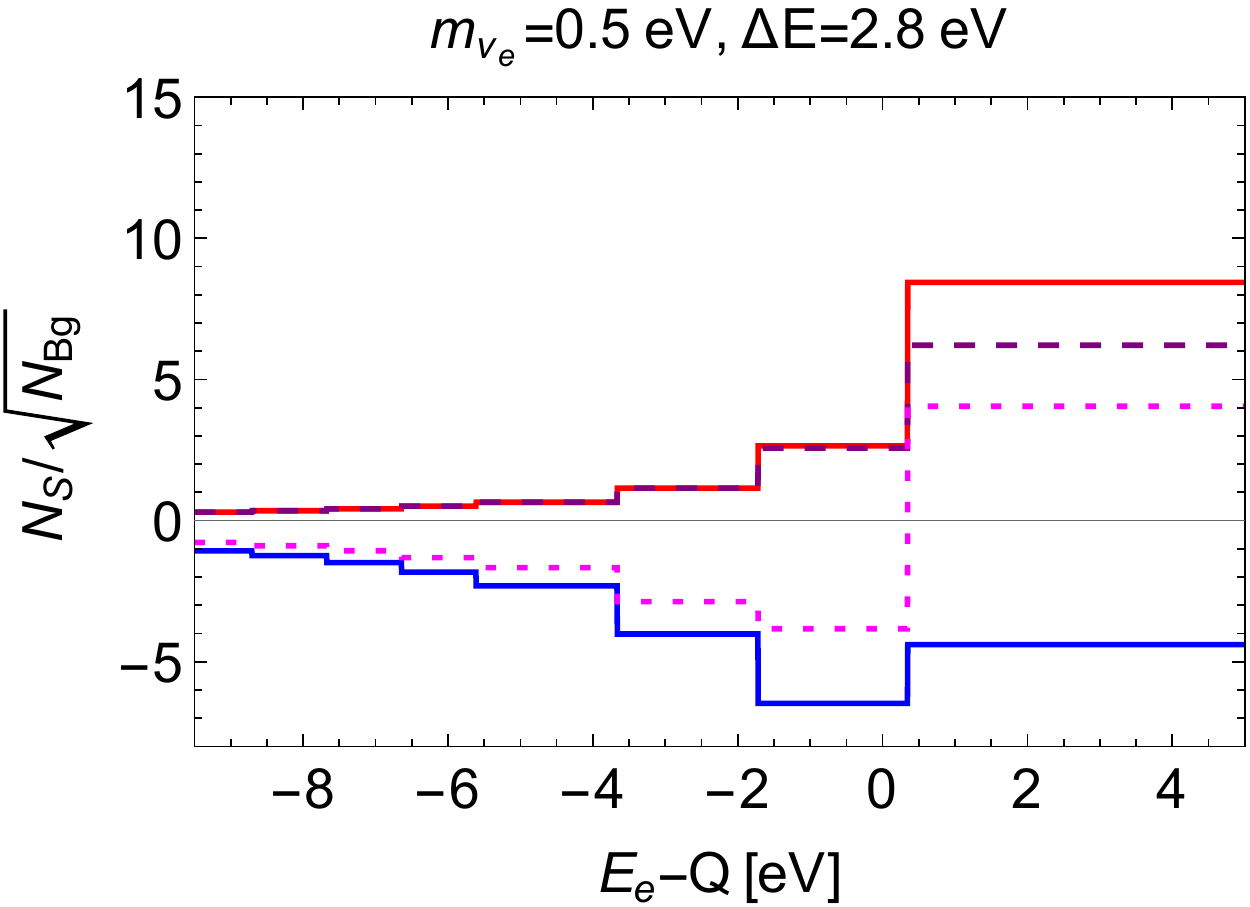}
    \caption{The number of expected signal events divided by square root of background events (standard tail of beta-decay) for different energy bins for massless neutrinos (left panel) and for massive neutrinos with $m_{\nu_e}=0.5$~eV (right panel). The figure is shown for a tritium mass of $13 \mu$g, an experimental observation time of $36$ hours per bin, and an energy resolution $\Delta E = 2.8$~eV (parameters of the KNM2 data set of KATRIN~\cite{KATRIN:2022kkv}). 
    We consider that the neutrino background forms a Fermi sphere in momentum space and show the case of  antineutrino overdensity $\bar{\eta} = 10^{11}$ (blue line),  neutrino overdensity $\eta = 10^{11}$ (red line), and the case of equal  neutrino and antineutrino overdensity $\eta = \bar{\eta} = 10^{11}$ (magenta dotted line). For comparison, we also show the assumption adopted in~\cite{KATRIN:2022kkv} that all neutrino momenta are negligibly small (purple dashed line) for $\eta = 10^{11}$.}
    \label{fig:energy-bins} 
\end{figure}

The  experimentally important process is neutrino capture (signal),
\begin{equation}
    \text{T} + \nu_e \to \,  ^3{\rm He} + e^-,
\end{equation}
for which the rate depends on the local \textit{neutrino} density $n_\nu = \eta n_0$, see Section~\ref{sec:capture}.
It is enabled beta-decay instability of the target (background),
\begin{equation}
    \text{T} \to  \, ^3{\rm He} + e^- + \bar{\nu}_e.
\end{equation}
Importantly, when considering large local overdensities, the endpoint region of the beta-spectrum becomes affected by the presence  of an \textit{antineutrino} background with number density $ n_{\bar\nu} =\bar \eta n_0 $~\cite{PhysRev.128.1457}.

In the following we demonstrate the modifications by adopting a value of $\eta = 10^{11} $ (and $\bar \eta = 10^{11}$) in line with the published work by the KATRIN collaboration~\cite{KATRIN:2022kkv}. We assume a Fermi gas at zero temperature,~i.e., all levels up to $E_F$ are filled.\footnote{We call Fermi energy $E_F = \sqrt{k_F^2 + m_{\nu_e}^2} - m_{\nu_e}$, where $k_F$ is the Fermi momentum.} For the capture reaction, such phase space distribution leads to a monotonically rising trend of the electron kinetic energy spectrum from $Q+m_{\nu_e}$ until $Q+m_{\nu_e}+E_F$ where $Q=18.56$~keV is the beta endpoint energy. In turn, Pauli blocking prohibits the release of an anti-electron neutrino with energy smaller than~$E_F$.%
\footnote{We note in passing that although we consider large neutrino asymmetries below, it does not modify the dispersion relation of free neutrinos. Parametrically, the energy shift is of  order $\Delta E \sim G_F n_0 (\eta - \bar\eta) \sim 10^{-25}\ {\rm eV}  $ for $\eta = 10^{11}$ and hence minute.}
This cuts off the beta spectrum for electron kinetic energies in the interval $[Q-m_{\nu_e}-E_F,Q-m_{\nu_e}]$. 
In Fig.~\ref{fig:peak_valley} we show examples of a signal from neutrino capture (red line) and modification of a beta-spectrum because of an antineutrino background (blue line) for two different neutrino masses. Details on the beta-spectrum and neutrino-capture rates are given in Appendix~\ref{app:all-rates}. 

One should keep in mind that, in  general, $\eta$ and $\bar{\eta}$ are independent parameters.  We  consider three different scenarios below:
\begin{enumerate}
    \item Presence of large neutrino background with overdensity $\eta$, but no significant antineutrino background, $\bar \eta \ll \eta$
    \item Presence of large antineutrino background with overdensity $\bar{\eta}$, but no significant neutrino background, $\bar \eta \gg \eta$;
    \item Presence of both neutrino and antineutrino background. For simplicity we assume the same overdensity $\eta = \bar{\eta}$ in this case.%
    \footnote{Matter interference effects induce an asymmetry close to the Earth's surface even when starting from a symmetric situation~\cite{Arvanitaki:2022oby,Arvanitaki:2023fij}.}
\end{enumerate}

\subsection{Toy-model estimate}

Let us consider an experimental setup with the energy resolution $\Delta E$. To model the finite experimental resolution we follow~\cite{PTOLEMY:2019hkd} and use Gaussian smearing with full-width-at-half-maximum (FWHM) $\Delta E$ to obtain the observed rate,
\begin{equation}
    \frac{d\widetilde{\Gamma}}{dE_{e}}(E_e) = \frac{1}{\sqrt{2\pi} (\Delta E/\sqrt{8 \ln 2})} \int \frac{d \Gamma}{dE_{e}}(E') \exp \left( - \frac{(E'-E_e)^2}{2 (\Delta E/\sqrt{8 \ln 2})^2} \right) dE'.
    \label{eq:smoothing}
\end{equation}
The demand on the required energy resolution to resolve the capture peak(s) from the beta-decay endpoint spectrum depends critically on the value of the neutrino mass(es). The current limit on the sum of the neutrino masses from cosmological structure formation, $\sum m_{\nu_i} < 0.12\ {\rm eV}\ (95\ \%~{\rm C.L.})$~\cite{Planck:2018vyg}, implies a maximal value of $30(16)$~meV for the lightest neutrino mass for normal (inverted) mass ordering when taking into account the solar and atmospheric mass splittings~\cite{ParticleDataGroup:2022pth}. This constitutes a challenging~\cite{PTOLEMY:2022ldz} requirement of $\Delta E = O(10)$~meV for C$\nu$B detection.%
\footnote{For a scenario where the cosmological neutrino mass bound is relaxed to eV-scale, see~\cite{Alvey:2021xmq}.}
The current KATRIN limit on the effective neutrino mass defined through $m_\beta^2 \equiv \sum m_{\nu_i}^2 |U_{ei}|^2$ stands at $m_\beta \leq  0.8\ {\rm eV}\ (90\%\ {\rm C.L.}) $~\cite{KATRIN:2021uub} with eV-scale energy resolution.

In Fig.~\ref{fig:smearing} we present the expected  differential rates in electron kinetic energy for  neutrino capture  (red line) and beta-decay (blue line). For comparison,  we also show the signal for  neutrino capture if the momentum distribution of neutrino were given by the Dirac delta function (purple dashed line), as it was assumed by KATRIN in~\cite{KATRIN:2022kkv}. We see that one should not neglect the effects of Pauli blocking, as it produces a sizable difference in differential rates for $\Delta E \sim E_F$ or better.

Adopting the energy resolution $\Delta E = 2.8$~eV we may directly compare the differences between the KATRIN result which neglects the neutrino phase space distribution and our results for various cases. Following~\cite{KATRIN:2022kkv}, we assume that the experiment counts the number of electrons with energy larger than some minimal value $E_{e,\min}$ and adopt the same energy bins (values of the minimal energies) as in the KNM2 run~\cite{KATRIN:2022kkv}. With a total mass of the tritium target of $m_{\text{T}} = 13~\mu$g we calculate the number of events in each energy bin within a window of $36$~h of exposure.  

The resulting signal-to-noise ratios are shown in Fig.~\ref{fig:energy-bins}.\footnote{We call ``noise'' events from $\beta$-electrons and  ``signal'' the difference between the number of events in the presence of a neutrino background and the case of its absence.} From this we see that the KATRIN experiment is sensitive to all three scenarios. The hardest scenario for exclusion is when both neutrino and antineutrino backgrounds are present, but even in this case the sensitivity of the KATRIN experiment drops only by a factor of few. Also, it is worth mentioning that the physically motivated Fermi-sphere distribution of neutrino (red line) is easier to detect than the unphysical Dirac delta-function distribution (purple dashed line). The difference between the two momentum distributions is stronger for low neutrino mass.

\section{Summary and conclusions}
\label{sec:conclusions}

Motivated by recent experimental results of the KATRIN experiment~\cite{KATRIN:2022kkv} which placed the limit $\eta \lesssim 10^{11}$, in this work we explore the viability of  such large overdensities of the low-energy neutrino background with respect to the standard cosmological C$\nu$B prediction. We show that the Pauli exclusion principle alone puts severe restrictions on the possible values of~$\eta$, even without detailed information about its formation mechanism or the present neutrino momentum distribution.

We find that both, the cosmological average value and any generated density (globally) within the Milky Way is restricted to $\eta \lesssim 10^4$. This bound is derived under the most accommodating assumptions but proposals exist that may saturate that number. An alternative avenue exists in the consideration of bound neutrino clusters, where, under the influence of a new force, $\eta \lesssim 10^7$ appears attainable. When one is willing to entertain the hypothesis of Solar System local sources, a local neutrino overdensity of $\eta \sim 10^{11}$ remains a theoretical possibility. A summary of the various studied origins is given in Fig.~\ref{fig:max_density}.

We furthermore discuss experimental signatures in the presence of a large neutrino/anti\-neutrino background. We show that Pauli blocking effects cannot be ignored, invalidating the assumptions on the electron neutrino distribution function made in~\cite{KATRIN:2022kkv}. Although the overall sensitivity to $\eta$ remains comparable, the expected event shape is altered.
If an overdensity in anti-electron neutrinos is present, $\bar \eta \gg 1$, the endpoint of the beta-spectrum is additionally modified. 
Those signatures are observable if the energy resolution of the experiment $\Delta E$ is in the ballpark of the Fermi energy $E_F$. For the case of KATRIN, this condition is  indeed fulfilled, $\Delta E \sim E_F \sim 1~{\rm eV}$, with their present sensitivity on~$\eta$. 

An improved sensitivity to more palatable values of $\eta, \bar \eta \lesssim 10^4$ of course requires significant experimental advances; we recall that a detection of the standard \CNB\ demands sensitivity to $\eta \lesssim 10$ when taking into account local clustering factors. 
Since $k_F \sim \eta^{1/3}$ we find that for $\eta, \bar \eta \sim 10^4 $, observation of the  imprints of Pauli blocking requires an energy resolution of $\Delta E \sim 10\ {\rm meV}$.%
\footnote{A detailed analysis of this case requires going beyond the effective ``one-neutrino'' picture used in this work since $\Delta E$ is at the level of the atmospheric neutrino mass splitting $\sqrt{|\Delta m_{\rm atm}^2|}\simeq 50$~meV.} It is clear though that such an advance in neutrino capture experiments offers the prospect of observing a spectacular signature of an otherwise most elusive dark radiation component in our Universe and provides an important intermittent physics target in the ultimate quest for a direct detection of the~C$\nu$B.

\paragraph{Acknowledgements} This work was supported by the Research Network Quantum Aspects of Spacetime (TURIS). KB is partly funded by the INFN PD51 INDARK grant.
AB is supported by the European Research Council (ERC) Advanced Grant ``NuBSM'' (694896). AS is supported  by the Kavli Institute for Cosmological Physics at the University of Chicago through an endowment from the Kavli Foundation and its founder Fred Kavli. This work has been supported by the Fermi Research Alliance, LLC under Contract No. DE-AC02-07CH11359 with the U.S. Department of Energy, Office of High Energy Physics. Funded/Co-funded by the European Union (ERC, NLO-DM, 101044443). Views and opinions expressed are however those of the author(s) only and do not necessarily reflect those of the European Union or the European Research Council. Neither the European Union nor the granting authority can be held responsible for them.

\appendix

\section{Beta spectrum and neutrino capture for tritium}
\label{app:all-rates}

For the differential beta-spectrum we modified the formula from~\cite{Masood:2007rc,PTOLEMY:2019hkd} taking into account that in the presence of antineutrinos that form a Fermi sphere with Fermi momentum $k_F$, some decay channels are forbidden by Pauli blocking principle; we simplified the formula to the case of only one neutrino flavor. The resulting differential beta-decay rate is given by,
\begin{equation}
    \frac{d\Gamma_\beta}{d E_e} = \frac{(\sigma v)_0}{\pi^2} N_T H(E_e, m_{\nu_e}) \Theta(Q- m_{\nu_e} - E_F - E_e),
\end{equation}
with
\begin{align}
    H(E_e, m_{\nu_e}) &= \frac{1 - m_e^2/ (E_e m_{\text{T}})}{\left( 1 - 2 E_e/m_{\text{T}} + m_e^2 /m_{\text{T}}^2 \right)^2}
    \sqrt{y\left(y + 2 m_{\nu_e}\frac{m_{^3\text{He}}}{m_{\text{T}}} \right)} \times \nonumber\\
    &\times 
    \left[ y + m_{\nu_e} \frac{m_{^3\text{He}} + m_{\nu_e}}{m_{\text{T}}} \right],
\end{align}
and where $y=Q-E_e-m_{\nu_e}$. The capture cross section $(\sigma v)_0$ is given in Sec.~\ref{sec:capture} and $E_F = \sqrt{k_F^2 + m_{\nu_e}^2} - m_{\nu_e}$, where $k_F$ depend on local neutrino density by Eq.~\eqref{eq:pauli}.

The differential neutrino capture can be obtained from~\eqref{eq:GammanuDM} together with energy conservation ($E_e = Q + E_{\nu_e}$),
\begin{equation}
    \frac{d \Gamma_{\text{capt}}}{d E_e} (E_e) = \frac{M_{\text{T}}}{m_{\text{T}}} (\sigma v)_0 \frac{dn_{\nu_e}}{dE_{\nu_e}}(E_e-Q),
    \label{eq:DM_rate}
\end{equation}
where the neutrino energy distribution is 
\begin{equation}
    \frac{d n_{{\nu_e}}}{d E_{\nu_e}}(E_{\nu_e}) \simeq \frac{g_{\nu_e}}{2 \pi^2} E_{\nu_e} \sqrt{E_{\nu_e}^2 - m_{\nu_e}^2} \Theta(m_{\nu_e} + E_F-E_{\nu_e})
    \label{eq:Fermi-energy-distribution}
\end{equation}
with $E_F = \sqrt{k_F^2 + m_{\nu_e}^2}-m_{\nu_e}$ and $g_{\nu_e} = 1$ for SM neutrinos; $k_F$ depends on the local neutrino density by Eq.~\eqref{eq:pauli}. Integrating Eq.~\eqref{eq:Fermi-energy-distribution} over neutrino energy yields~$n_{{\nu_e}} = \eta n_0$.

\bibliographystyle{JHEP}
\bibliography{refs.bib}

\end{document}